\documentclass[11pt,a4paper]{article}

\pdftrailerid{}

\usepackage[margin=2.5cm]{geometry}
\usepackage{amsmath,amssymb,amsfonts}
\usepackage[caption=false]{subfig}
\usepackage{graphicx}
\usepackage{dcolumn}
\usepackage{bm}
\usepackage{authblk}

\usepackage{siunitx}
\DeclareSIUnit\atomicmassunit{u}
\usepackage[version=4]{mhchem}
\usepackage{svg}
\usepackage{placeins}
\usepackage{enumitem}

\usepackage{tikz}
\usetikzlibrary{
  calc,
  patterns,
  backgrounds,
  arrows.meta
}

\usepackage{pgfplots}
\pgfplotsset{
  compat=1.18,
  table/col sep=comma,
}
\usepgfplotslibrary{
  units,
  groupplots,
  fillbetween
}

\definecolor{TUMBlue}        {RGB/cmyk}{  0,101,189 / 1.  ,0.43,0.  ,0.  }
\definecolor{TUMWhite}       {RGB/cmyk}{255,255,255 / 0.  ,0.  ,0.  ,0.  }
\definecolor{TUMBlack}       {RGB/cmyk}{  0,  0,  0 / 0.  ,0.  ,0.  ,1.  }

\definecolor{TUMBlueDark}    {RGB/cmyk}{  0, 82,147 / 1.  ,0.54,0.04,0.19}
\definecolor{TUMBlueDarker}  {RGB/cmyk}{  0, 51, 89 / 1.  ,0.57,0.12,0.7 }
\definecolor{TUMGrayDark}    {RGB/cmyk}{ 88, 88, 90 / 0.  ,0.  ,0.  ,0.8 }
\definecolor{TUMGray}        {RGB/cmyk}{156,157,159 / 0.  ,0.  ,0.  ,0.5 }
\definecolor{TUMGrayLight}   {RGB/cmyk}{217,218,219 / 0.  ,0.  ,0.  ,0.2 }

\definecolor{TUMGreen}       {RGB/cmyk}{162,173,  0 / 0.35,0.  ,1.  ,0.2 }
\definecolor{TUMOrange}      {RGB/cmyk}{227,114, 34 / 0.  ,0.65,0.95,0.  }
\definecolor{TUMIvory}       {RGB/cmyk}{218,215,203 / 0.03,0.04,0.14,0.08}
\definecolor{TUMBlueLight}   {RGB/cmyk}{100,160,200 / 0.65,0.19,0.01,0.04}
\definecolor{TUMBlueLighter} {RGB/cmyk}{152,198,234 / 0.42,0.09,0.  ,0.  }

\definecolor{TUMExtViolet}   {RGB/cmyk}{105,  8, 90 / 0.5 ,1.  ,0.  ,0.4 }
\definecolor{TUMExtNavy}     {RGB/cmyk}{ 15, 27, 95 / 1.  ,1.  ,0.  ,0.4 }
\definecolor{TUMExtTeal}     {RGB/cmyk}{  0,119,138 / 1.  ,0.03,0.3 ,0.3 }
\definecolor{TUMExtForest}   {RGB/cmyk}{  0,124, 48 / 1.  ,0.  ,1.  ,0.2 }
\definecolor{TUMExtLime}     {RGB/cmyk}{103,154, 29 / 0.6 ,0.  ,1.  ,0.2 }
\definecolor{TUMExtYellow}   {RGB/cmyk}{255,220,  0 / 0.  ,0.1 ,1.  ,0.  }
\definecolor{TUMExtGoldenrod}{RGB/cmyk}{249,186,  0 / 0.  ,0.3 ,1.  ,0.  }
\definecolor{TUMExtPumpkin}  {RGB/cmyk}{214, 76, 19 / 0.  ,0.8 ,1.  ,0.1 }
\definecolor{TUMExtRed}      {RGB/cmyk}{196,  7, 27 / 0.1 ,1.  ,1.  ,0.1 }
\definecolor{TUMExtMaroon}   {RGB/cmyk}{156, 13, 22 / 0.  ,1.  ,1.  ,0.4 }

\colorlet{black}      {TUMBlack}
\colorlet{blue}       {TUMBlue}
\colorlet{darkgray}   {TUMGrayDark}
\colorlet{gray}       {TUMGray}
\colorlet{green}      {TUMGreen}
\colorlet{lightgray}  {TUMGrayLight}
\colorlet{lime}       {TUMExtLime}
\colorlet{orange}     {TUMOrange}
\colorlet{red}        {TUMExtRed}
\colorlet{teal}       {TUMExtTeal}
\colorlet{violet}     {TUMExtViolet}
\colorlet{white}      {TUMWhite}
\colorlet{yellow}     {TUMExtYellow}
\colorlet{ForestGreen}{TUMExtForest}
\colorlet{Goldenrod}  {TUMExtGoldenrod}
\colorlet{Maroon}     {TUMExtMaroon}
\colorlet{NavyBlue}   {TUMExtNavy}
\colorlet{RedOrange}  {TUMExtPumpkin}

\pgfplotsset{
    colormap={parula5}{
        rgb255=(53,42,135)
        rgb255=(15,92,221)
        rgb255=(18,125,216)
        rgb255=(7,156,207)
        rgb255=(21,177,180)
        rgb255=(89,189,140)
        rgb255=(165,190,107)
        rgb255=(225,185,82)
        rgb255=(252,206,46)
        rgb255=(249,251,14)
    }
}

\title{Establishing the Magnetoelastic Origin of Spin-Wave Routing through Focused Ion Beam Patterning}

\author[1]{Felix Naunheimer\thanks{felix.naunheimer@tum.de}}
\author[1]{Johannes Greil}
\author[1]{Valentin Ahrens}

\author[2]{Levente Maucha}
\author[2]{\'Ad\'am Papp}
\author[2]{Gy\"orgy Csaba}
\author[1]{Markus Becherer}

\affil[1]{TUM School of Computation, Information and Technology, Technical University of Munich, Munich, Germany}
\affil[2]{Faculty of Information Technology and Bionics, Pázmány Péter Catholic University, Budapest, Hungary}

\date{\today}

\begin{document}
\maketitle

\begin{abstract}
Spin waves are promising information carriers for analog and wave-based computing, where functionality relies on compact and precisely engineered scattering landscapes. Focused ion beam (FIB) irradiation enables such control by locally tailoring the spin-wave dispersion in yttrium iron garnet (YIG). However, a non-monotonic dependence of the spin-wave wavelength on increasing ion dose hinders predictive landscape design.
Here, we present an experimentally validated framework that explains this non-monotonic spin-wave steering by linking phenomenological strain-induced anisotropy to its magnetoelastic origin. Irradiation-induced lattice dislocations drive elastic and plastic deformation, which evolve into partial amorphization, each stage contributing distinctly to the dispersion behavior.
We combine post-irradiation wet-chemical etching and atomic force microscopy (AFM) to quantify thickness changes, and track the dispersion in etched regions using time-resolved magneto-optical Kerr effect (trMOKE) microscopy. Fitting the data to the Kalinikos–Slavin formalism with an added effective magnetoelastic field isolates contributions from elastic and plastic deformation.
Validation is achieved by mapping the deformation evolution onto a three-phase scenario based on SRIM simulations, reproducing the extracted field trends, and by consistent strain tensor and micromagnetic analyses.
These results establish a physical basis for FIB-engineered graded-index (GRIN) spin-wave landscapes and magnetoelastically programmable magnonic devices.
\end{abstract}

\section{Introduction}
Analog computing is attracting renewed attention in the era of AI, particularly in application domains where CMOS technology encounters fundamental limitations. Rather than aiming to replace CMOS in general-purpose digital logic, alternative wave-based approaches focus on specialized, high-frequency, low-power, and non-Boolean computing tasks \cite{csaba2017perspectives}.

In this context, spin waves, collective excitations of a spin system that propagate precessional motion in a wave-like manner, are emerging as promising information carriers. Their quasiparticles, called magnons, naturally allow for operation in the gigahertz regime \cite{mahmoud2020introduction}, making them intrinsically suited for microwave and RF signal-processing applications where low-power CMOS implementations remain challenging. Moreover, their wavelengths at gigahertz frequencies can be orders of magnitude shorter than those of electromagnetic waves, reaching the sub-micrometer scale. This wavelength compression enables highly compact analog processing elements and interference-based computing structures that cannot be efficiently realized with conventional RF electronics \cite{csaba2017perspectives, mahmoud2020introduction}. 

Crucially, many spin-wave functionalities are governed by the local dispersion relation: for a fixed excitation frequency, a controlled local change in the dispersion translates into a controlled change of the local wave vector and wavelength.
Despite their nonlinear dispersion, spin waves admit close analogies to optics and support quasi-optical elements \cite{csaba2014spin}. Spatially tailoring the effective magnetization, and thus the local spin-wave dispersion, allows the realization of graded refractive-index (GRIN) profiles for precise routing of spin waves \cite{davies2015graded}. In magnonic implementations, the term “GRIN” therefore refers to an engineered spatial profile of the dispersion (or equivalently the local wavelength at a given frequency), which governs the bending of wave fronts and the focusing or defocusing of spin-waves. Such quasi-optical elements include lenses and related routing geometries \cite{kiechle2023spin}. More complex scattering landscapes, e.g., those obtained by inverse design, have also been explored, but they ultimately rely on the same physical requirement: a reproducible, quantitative handle on how local magnetic modifications translate into local dispersion engineering \cite{wang2021inverse,papp2021nanoscale}.

Among available nanofabrication approaches, focused ion beam (FIB) irradiation provides an attractive route to implement such dispersion landscapes in yttrium iron garnet (YIG), owing to its high spatial resolution and the tunability of the interaction depth via acceleration voltage and ion species. Ion-based magnetic modification has been used in several magnetic-device contexts, including field-coupled magnetic logic \cite{Breitkreutz2012b, Becherer2016a} and the controlled nucleation and motion of skyrmions \cite{ahrens2022skyrmion,ahrens2023skyrmions}. In YIG specifically, localized ion implantation has been shown to enable dispersion-tunable, low-loss spin-wave waveguides \cite{bensmann2025dispersion}. In the context of spin-wave routing, our group previously demonstrated that $30~\si{\kilo\electronvolt}$ $\mathrm{Ga}^{+}$ FIB irradiation enables steering of spin-wave wavefronts, which was attributed to a local modulation of the effective magnetization \cite{kiechle2023spin, kiechle2022experimental, greil2025effect}.

Despite these device-level demonstrations, the crystallographic and microstructural origin of FIB-induced dispersion shifts in YIG thin films remains insufficiently established. The aforementioned effective magnetization modulation, and the associated wavelength changes, have so far been described using an effective, phenomenological strain-induced anisotropy. While ion implantation is known to enhance strain-induced anisotropy in garnet films \cite{eschenfelder2012magnetic, wolfe1971modification, jouve1979specific, mada1979fmr, ruane2018controlling}, such approximations do not fully capture key experimental trends, in particular the non-monotonic (approximately parabolic) evolution of the effective refractive-index-like response reported in Refs. \cite{kiechle2023spin, greil2025effect}. This motivates a model that explicitly accounts for irradiation-driven structural evolution, including defect generation, strain redistribution, and partial amorphization \cite{fodchuk2022effect,was2007fundamentals}.

In this work, we propose a three-phase scenario describing the structural evolution of ion-implanted YIG and the resulting non-monotonic behavior reported in Refs. \cite{kiechle2023spin, greil2025effect}. While these studies primarily address application-relevant aspects, our work elucidates the physical origin of this non-monotonicity by identifying the underlying microscopic mechanisms and organizing them into three distinct regimes:
\begin{enumerate}[label=(\Roman*)]
\item Strain accumulation driven by lattice dislocations formed in radiation-induced collision cascades;
\item Strain relaxation through the migration of dislocations toward regions of lower strain; and
\item Near-surface amorphization, accompanied by continued strain accumulation in deeper layers governed by the mechanisms active in phases (I) and (II).
\end{enumerate}

We test this scenario using a self-consistent experimental–computational pipeline. Following FIB irradiation, wet-chemical etching is employed to remove the near-surface amorphous material formed in phase (III). The resulting local thickness reduction is quantified by atomic force microscopy (AFM) and used as a fixed geometric constraint in the subsequent analysis. Spin-wave dispersion relations are then measured in the etched regions using time-resolved magneto-optical Kerr effect (trMOKE) microscopy and analyzed by fitting the Kalinikos–Slavin formalism \cite{kalinikos1986theory}, extended by an explicit magnetoelastic field term to account for strain accumulation and relaxation associated with phases (I) and (II).

To justify the use of the magnetoelastic field as a representative parameter for irradiation-induced strain accumulation and relaxation, and the use of the local thickness reduction as an indicator of partial amorphization, we introduce an analogous three-phase scenario based on SRIM Monte Carlo simulations \cite{ziegler2010srim}. As a final consistency check, strain tensor components derived independently from the fitted magnetoelastic field and from the modeled strain evolution are implemented in micromagnetic simulations \cite{vansteenkiste2014design} to reproduce the experimentally observed trends in the dispersion of FIB-steered spin waves.

\section{Methods}
\label{methods}
\begin{figure}[t!]
    \centering
  \includegraphics[width=\linewidth]{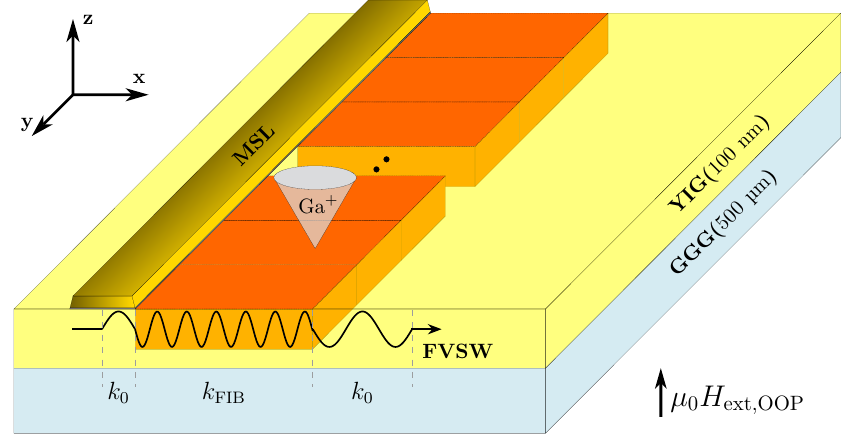}
    \caption{Schematic overview of a $100~\si{\nano\meter}$ YIG thin film deposited by RF-Sputtering on a $500~\si{\micro\meter}$ GGG substrate. Spin wave excitation is provided by a $2~\si{\micro\meter}$ wide, $110~\si{\nano\meter}$ thick Ti($10~\si{\nano\meter}$)/Au($100~\si{\nano\meter}$) MSL. Adjacent to the antenna, squares of $50\times50~\si{\micro\meter\squared}$ (red), were irradiated with $\mathrm{Ga}^{+}$ ion doses ranging from $2$ to $60\times10^{12}~\si{ions\per\centi\meter\squared}$ in steps of $2\times10^{12}~\si{ions\per\centi\meter\squared}$. In the FVSW configuration, spin waves are excited at a $k_0$ wave vector directly beneath the antenna. Upon entering the irradiated regions, the wave vector changes to $k_{\mathrm{FIB}}$, and returns to $k_0$ after leaving the region.}
    \label{fig:scatch}
\end{figure}

To investigate the impact of $\mathrm{Ga}^{+}$ ion irradiation on propagating spin waves, the experimental setup shown schematically in Fig.~\ref{fig:scatch} (not to scale) was employed. 

A YIG thin film with a thickness of $t = 100~\si{\nano\meter}$ was deposited by RF sputtering onto a $500~\si{\micro\meter}$ GGG substrate and subsequently annealed in an oxidation furnace to obtain the desired crystalline structure.
For spin-wave excitation, a Ti($10~\si{\nano\meter}$)/Au($100~\si{\nano\meter}$) microstrip line (MSL) with a width of $2~\si{\micro\meter}$ was deposited by e-beam evaporation.
Adjacent to the MSL, squares of $50 \times 50~\si{\micro\meter\squared}$ were irradiated by direct FIB writing at varying $\mathrm{Ga}^{+}$ ion doses ranging from $2$ to $60 \times 10^{12}~\si{ions\per\centi\meter\squared}$ in steps of $2 \times 10^{12}~\si{ions\per\centi\meter\squared}$.
The irradiated regions were written at acceleration voltages of $30~\si{\kilo\electronvolt}$, $16~\si{\kilo\electronvolt}$, and $8~\si{\kilo\electronvolt}$ to obtain different penetration depths of the ions.
To experimentally verify the reduction in effective film thickness caused by partial surface amorphization, as reported in Ref. \cite{fodchuk2022effect}, we introduced an additional wet-chemical etching step. The film was etched for $120~\si{\second}$, and the resulting stepwise changes in etching depth across the irradiated regions were quantified by atomic force microscopy (AFM).

Spin-wave propagation in the irradiated regions was investigated in the forward-volume spin-wave (FVSW) configuration.
Spin waves were excited at a wavenumber $k_{0}$ in the unirradiated region beneath the stripline and coupled into the irradiated area, where the wavenumber changed to $k_{\mathrm{FIB}}$, as illustrated in Fig.~\ref{fig:scatch}.
Upon exiting the irradiated square, the wavenumber reverted to its original value $k_{0}$.
The experiment was conducted using trMOKE microscopy, allowing us to analyze spin-wave propagation across $30$ distinct implantation regions as well as in the pristine film at the intrinsic wavenumber $k_{0}$.
This approach enabled us to track the evolution of the spin-wave dispersion as a function of implantation dose and to extract key parameters through fits to the Kalinikos–Slavin model.
The measurements were performed at spin-wave excitation frequencies ranging from $2.285~\si{\giga\hertz}$ to $2.33~\si{\giga\hertz}$ in steps of $5~\si{\mega\hertz}$. An input power of $8~\si{\decibel m}$ was chosen to keep the measured spin-wave signal just above the noise floor, thereby minimizing nonlinear contributions that arise at higher input powers, as reported in Ref. \cite{ogawa2025quantitative}. An external magnetic field of $\mu_{0}H_{\mathrm{ext,OOP}} = 250~\si{\milli\tesla}$ was applied in the out-of-plane (OOP) direction.

To model the depth-dependent damage profile induced by $\mathrm{Ga}^{+}$ ion irradiation, SRIM Monte Carlo simulations were performed. The simulations calculate the energy loss of individual ions within the crystal lattice and generate displacement cascades that lead to the formation of vacancies, interstitials, and lattice disorder throughout the film thickness.

The ion species was defined as $\mathrm{Ga}^{+}$ with a mass of approximately $70~\si{\atomicmassunit}$. Acceleration voltages of $30~\si{\kilo\electronvolt}$, $16~\si{\kilo\electronvolt}$, and $8~\si{\kilo\electronvolt}$ were used to study the energy dependence of the damage profile. The simulated YIG layer had a density of $5.17~\si{g\per\centi\meter\cubed}$ and a thickness of $100~\si{\nano\meter}$, with a stoichiometry of \ce{Y3Fe5O12}. Displacement energies of $E_{\mathrm{Y}} = 66~\si{\eV}$, $E_{\mathrm{Fe}} = 56~\si{\eV}$, and $E_{\mathrm{O}} = 40~\si{\eV}$ were used \cite{ubizskii2000displacement}. To minimize channeling effects \cite{fodchuk2022effect}, the ion beam was tilted by $7^{\circ}$ with respect to the surface normal.

From the SRIM output, the depth-dependent damage profile expressed in units of $(\si{atoms\per\centi\meter\cubed})/(\si{ions\per\centi\meter\squared})$, is given. By multiplying this profile with the experimentally applied ion dose, the depth-dependent damage model $\mathrm{DM}_{\mathrm{FIB}}$ was constructed, yielding the local damage density as a function of depth in units of $\si{atoms\per\centi\meter\cubed}$.

Micromagnetic simulations were performed to substantiate the magnetoelastic origin of the experimentally observed dispersion shifts. To isolate this effect, the simulations were restricted to the irradiated regions only, modeled as homogeneous squares with uniformly distributed material parameters across the entire simulation grid. Consequently, no spatial variations, interfacial regions, or gradients between irradiated and non-irradiated areas were considered. 

The saturation magnetization was set to $M_{\mathrm{s}} = 130~\si{kA\per\meter}$, as determined from on-chip ferromagnetic resonance (FMR) measurements. A magnetoelastic coupling constant of $B_{1} = 3.48 \times 10^{5}~\si{J\per\meter\cubed}$, taken from bulk material values \cite{smith1963magnetostriction}, was assumed for the sputtered YIG films, together with a Poisson ratio of  $\nu = 0.29$ \cite{bhoi2018stress}. Magnetoelastic constants in thin films can deviate from bulk values due to growth- and strain-related effects, as reported for single-crystal YIG films in Ref. \cite{an2023optimizing}, which may introduce quantitative uncertainties in the simulations. However, the present approach is intentionally focused on providing a simple and accessible physical description of the dominant effects. 

\section{Results}
\label{results}
\subsection{Spin-Wave Scattering in Ion-Implanted Thin Films}
\label{observation}
\begin{figure}[b!]
  \centering
  \includegraphics[width=\linewidth]{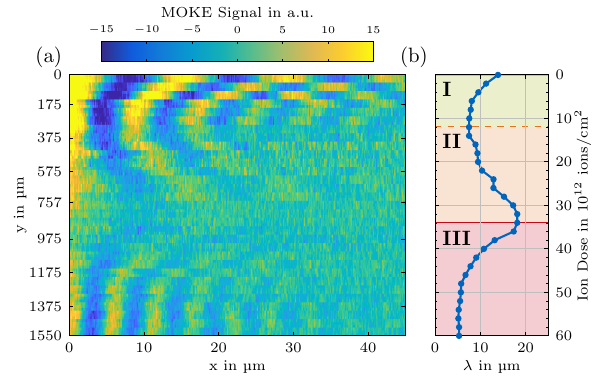}
  \caption{trMOKE measurement of a $t = 100~\si{\nano\meter}$ wet-chemically etched YIG film at $2.305~\si{\giga\hertz}$ and $8~\si{\mathrm{dBm}}$ input power with $\mu_{0}H_{\mathrm{ext,OOP}} = 250~\si{\milli\tesla}$.
(a) trMOKE image of coherently excited spin waves across $30$ distinct $\mathrm{Ga}^{+}$-implanted regions (irradiated at $30~\si{\kilo\electronvolt}$) over a propagation distance of $45~\si{\micro\meter}$.
(b) Line-wise Fourier transformations of the spin-wave profiles in (a) reveal wavelength shifts as a function of ion dose. Three prominent wavelength regims are observed: I between the unimplanted reference ($0 \times 10^{12}~\si{ions\per\centi\meter\squared}$) and the first turning point at $12 \times 10^{12}~\si{ions\per\centi\meter\squared}$; II between $12 \times 10^{12}~\si{ions\per\centi\meter\squared}$ and $34 \times 10^{12}~\si{ions\per\centi\meter\squared}$; and III extending from $34 \times 10^{12}~\si{ions\per\centi\meter\squared}$ to a saturation regime near $60 \times 10^{12}~\si{ions\per\centi\meter\squared}$.}
  \label{fig:dosemap}
\end{figure}

As previously demonstrated by our group \cite{kiechle2023spin}, $\mathrm{Ga}^{+}$ ion irradiation via direct FIB writing on a YIG thin film modifies the magnetic properties and, consequently, the spin-wave wavelength. To correlate the wavelength modification with the corresponding ion dose, we constructed a characteristic dosemap comprising 
$30$ irradiation regions, as shown in Fig.~\ref{fig:dosemap}. The dosemap captures the spin-wave scattering across regions implanted at an ion acceleration of $30~\si{\kilo\electronvolt}$, with doses ranging from $2$–$60\times10^{12}~\si{ions\per\centi\meter\squared}$. The plane-wave fronts measured by trMOKE at $2.305~\si{\giga\hertz}$ and $8~\si{\mathrm{dBm}}$ input power with $\mu_{0}H_{\mathrm{ext,OOP}} = 250~\si{\milli\tesla}$ (Fig.~\ref{fig:dosemap}(a)) were analyzed via a line-wise Fourier transformation along the $x$-direction to extract the carrier wavelength $\lambda$ (Fig.~\ref{fig:dosemap}(b)). 
This analysis reveals two clear turning points at doses of $12\times10^{12}~\si{ions\per\centi\meter\squared}$ and $34\times10^{12}~\si{ions\per\centi\meter\squared}$, which divide the dose-dependent response into three monotonic wavelength regimes. These regimes are characterized by decreasing (I), increasing (II), and again decreasing (III) wavelengths.

In Ref.~\cite{kiechle2023spin}, wavelength regimes I and II were attributed to strain-induced magnetic anisotropy generated by ion irradiation, which modifies the effective magnetization $M_{\mathrm{eff}}$ and thereby shifts the spin-wave dispersion to shorter or longer wavelengths. While this strain-induced anisotropy captures monotonic trends of the wavelength evolution, it does not account for the pronounced turning points that define the three regimes in Fig.~\ref{fig:dosemap}(b). As we show in the following sections, this non-monotonicity arise from the nucleation and motion of irradiation-induced crystalline dislocations within the YIG lattice, which establish the strain responsible for the observed anisotropy.

\subsection{Three-Phase Scenario for Irradiation-Induced Deformations}
\label{interpretation}
\begin{figure}[h!]
  \centering
  \includegraphics[width=\linewidth]{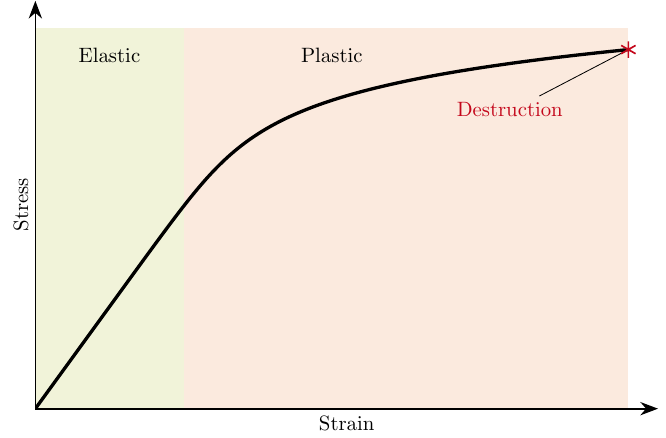}
  \caption{Schematic stress–strain curve of a crystal under continuous irradiation, illustrating elastic and plastic deformation up to the breakdown of the crystalline structure.}
  \label{fig:deformation}
\end{figure}

To evaluate the stability and mechanical strength of a material, stress testing is commonly employed to characterize its stress–strain response.
The characteristic stress response of a material under linearly increasing strain is represented by its stress–strain curve (Fig.~\ref{fig:deformation}).
This curve illustrates the transition from the reversible elastic deformation regime to the irreversible plastic deformation regime, ultimately culminating in a destructive deformation regime of the specimen.

\begin{figure}[h!]
    \centering
  \includegraphics[width=\linewidth]{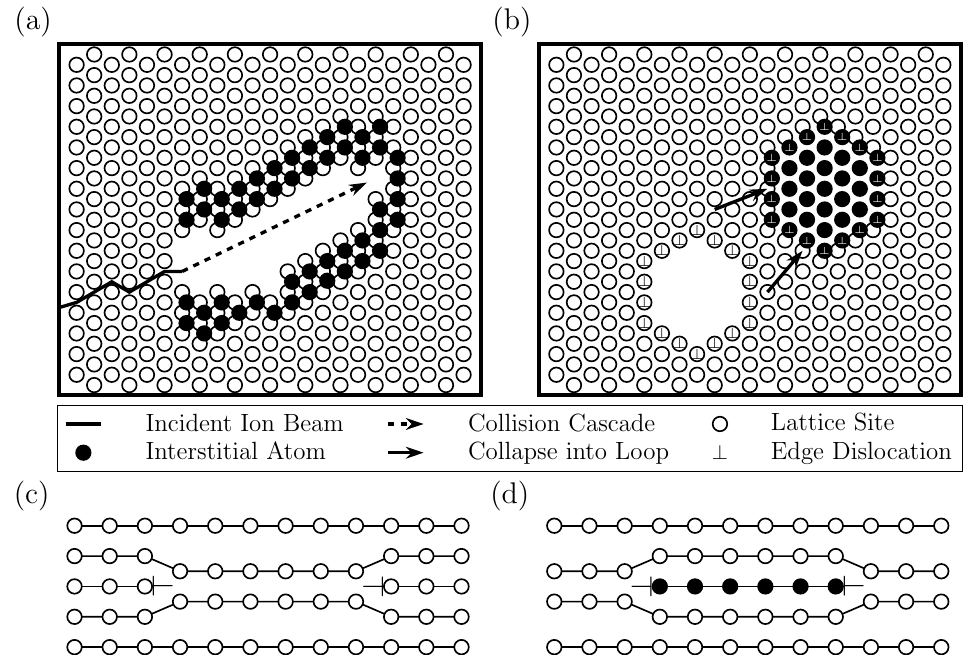}
    \caption{Illustration of the dislocation formation.
    (a) An incident ion beam initiates collision cascades that displace atoms from their lattice sites, creating a vacancy-rich core surrounded by interstitial atoms that aggregate into a defect cluster.
    (b) With increasing defect concentration, the cluster collapses into an energetically favored edge dislocation ($\perp$) loop. (c) A vacancy-type dislocation loop corresponds to a missing atomic plane, causing the surrounding lattice to bend inward toward the vacant lattice sites.
    (d) An interstitial-type dislocation loop corresponds to an additional atomic plane, pushing the lattice outward from the additional plane.}
    \label{fig:nucleation}
\end{figure}

As this regimes originate at the atomic scale, the concept of a stress–strain curve can be extended to describe the effects of ion irradiation on crystalline materials. In this analogy, ion irradiation acts as an effective pressure applied to the lattice, gradually altering its internal structure. When accelerated ions penetrate the crystal, they initiate collision cascades that displace atoms from their lattice sites, producing a vacancy-rich core surrounded by an interstitial atom shell. With increasing irradiation dose, these defect regions grow and aggregate into compact clusters (Fig.~\ref{fig:nucleation}(a)). These clusters subsequently collapse into energetically favored edge dislocation ($\perp$) loops, as shown in Fig.~\ref{fig:nucleation}(b), which bend the surrounding lattice and induce strain \cite{was2007fundamentals}.
Figures~\ref{fig:nucleation}(c)-\ref{fig:nucleation}(d) illustrate the lattice bending produced by these dislocation loops. A vacancy-type loop generates an edge dislocation with step-like distortions in the adjacent lattice planes (Fig.~\ref{fig:nucleation}(c)), while Fig.~\ref{fig:nucleation}(d) shows the corresponding structure formed by an interstitial loop. As the irradiation dose increases, the density of such loops grows accordingly, which increases the magnitude of induced strain.

\begin{figure}[t!]
  \centering
  \includegraphics[width=\linewidth]{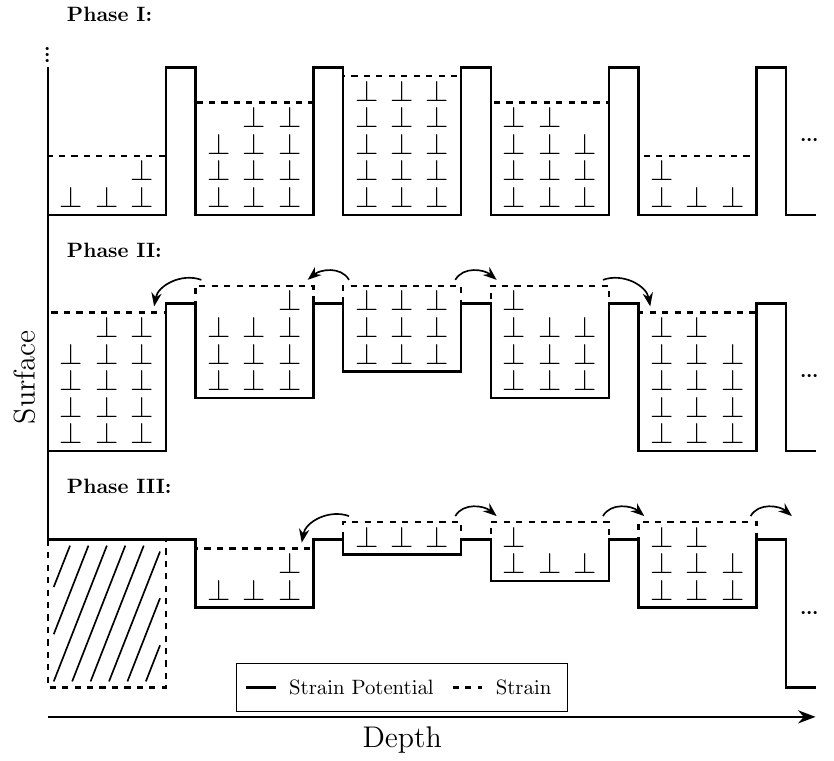}
  \caption{Conceptual illustration of strain potential wells that translate dislocation dynamics into a three-phase scenario of irradiation-induced deformations. Phase I: Dislocations ($\perp$) nucleate and accumulate within the strain well at the ion-damage source, progressively filling it in a process known as source hardening. Phase II: As the local strain exceeds the potential threshold, dislocations overcome the barrier and migrate into neighboring wells, initiating plastic deformation. Phase III: Migrated dislocations pile up against obstacles such as surfaces, defects, or grain boundaries, producing friction hardening and ultimately leading to partial amorphization~(//).}
  \label{fig:strain_wells}
\end{figure}

With the strain mechanism established, we turn to the stress-strain curve a crystalline material undergoes when exposed to ion irradiation, as illustrated in Fig.~\ref{fig:deformation}. To describe this curve, we map the elastic, plastic, and destructive deformation regimes onto three phases within a strain–potential landscape. This landscape is developed by orienting the description around the Peierls potential, which represents the periodic lattice energy barrier that must be overcome for a dislocation to move to an adjacent lattice row \cite{anderson2017theory}. The description further builds on established understanding of dislocation formation and nucleation under ion irradiation \cite{was2007fundamentals}. This strain-potential landscape can be viewed as a simplified, one-dimensional representation of the crystal lattice (Fig.~\ref{fig:strain_wells}). In this picture, the lattice is modeled as a sequence of potential wells distributed along the material depth, each separated by a barrier analogous to the Peierls potential. 
These barriers represent the strain that must be overcome for a dislocation to move from one lattice row to the next.

In the following, we define a three-phase deformation scenario that captures the evolution of dislocation behavior and strain accumulation under increasing ion irradiation.
In phase I, dislocations nucleate and begin to pile up around the ion-induced damage region. As the ion dose increases linearly, the number of dislocations rises proportionally, and the accumulated strain increases in a nearly linear manner, characteristic of an elastic response. In the strain-potential landscape, this corresponds to the potential well at the damage source gradually filling with dislocations, which is known as source hardening. 
In phase II, the increasing strain becomes sufficient for dislocations to overcome the potential barriers between neighboring wells. Once this threshold is exceeded, dislocations begin to move, interact, and rearrange, marking the onset of plastic deformation. In Fig.~\ref{fig:strain_wells}, this corresponds to dislocations hopping from one well to the next as the local stress surpasses the Peierls-like barrier. Importantly, dislocations preferentially migrate toward regions of lower strain, thereby partially relaxing the local strain while redistributing deformation throughout the lattice.
In phase III, the migrated dislocations accumulate once more, but now they pile up against obstacles such as the material surface, defects, or grain boundaries, which is known as friction hardening. Unlike in phase II, the locally increasing strain is no longer sufficient to overcome the potential barriers at these obstacles. As a result, dislocations become trapped, leading to a buildup of strain that cannot be relieved by further migration. In the strain-potential landscape, this corresponds to an effectively infinite barrier at the obstacle, producing a completely filled potential well that represents the maximum local strain the lattice can sustain. When this limit is reached, the crystalline order begins to break down, marking the onset of partial amorphization.

\subsection{Spin-Wave Dispersion in the Presence of Magnetoelastic Fields and Local Thickness Reduction}
\label{characterization}
To quantitatively link the three-phase scenario described above to the experimentally observed wavelength evolution in Sec.~\ref{observation}, we describe the effect of strain on spin-wave dynamics within the Landau–Lifshitz–Gilbert (LLG) formalism. The purpose of this section is to establish a direct correspondence between irradiation-driven strain evolution, experimentally accessible dispersion parameters, and the effective magnetic fields governing spin-wave propagation.

The strain state of a crystalline material is generally represented by the strain tensor $\boldsymbol{\varepsilon}(\vec{r})$. For brevity, the explicit spatial dependence is omitted here:
\begin{equation}
\begin{split}
\boldsymbol{\varepsilon} &=
\begin{bmatrix}
\varepsilon_{xx} & \varepsilon_{xy} & \varepsilon_{xz}\\
\varepsilon_{yx} & \varepsilon_{yy} & \varepsilon_{yz}\\
\varepsilon_{zx} & \varepsilon_{zy} & \varepsilon_{zz}
\end{bmatrix}.
\end{split}
\label{eq:strain_tensor}
\end{equation}
This tensor captures both the normal components $(\varepsilon_{xx}, \varepsilon_{yy}, \varepsilon_{zz})$ and the shear components $(\varepsilon_{yx}, \varepsilon_{zx}, \varepsilon_{zy})$ of the deformation. In micromagnetic simulations, strain enters the spin dynamics through magnetoelastic coupling, giving rise to an additional effective field contribution $\mathbf{H}_{\mathrm{mel}}$ \cite{vanderveken2021finite}.
Within the LLG equation, the magnetization dynamics are governed by the total effective field
\begin{equation}
\label{eq:eff_field}
    \mathbf{H}_{\mathrm{eff}}= \mathbf{H}_{\mathrm{eff},0} + \mathbf{H}_{\mathrm{mel}},
\end{equation}
where the magnetoelastic contribution $\mathbf{H}_{\mathrm{mel}}$ directly couples lattice deformations to the magnetic precession, in addition to the strain-free field components contained in $\mathbf{H}_{\mathrm{eff,0}}$.
The magnetoelastic contribution, $\mathbf{H}_{\mathrm{mel}}$, can be expressed by Eq.~(\ref{eq:strain_field}), which is commonly used in micromagnetic simulations~\cite{vanderveken2021finite}
\begin{equation}
  \mathbf{H}_{\mathrm{mel}}=-\frac{2}{\mu_0M_{\mathrm{s}}}
    \begin{bmatrix}
        B_1\varepsilon_{xx}m_x+B_2(\varepsilon_{xy}m_y+\varepsilon_{zx}m_z)\\
        B_1\varepsilon_{yy}m_y+B_2(\varepsilon_{xy}m_x+\varepsilon_{yz}m_z)\\
        B_1\varepsilon_{zz}m_z+B_2(\varepsilon_{zx}m_x+\varepsilon_{yz}m_y)
    \end{bmatrix}
    ,
  \label{eq:strain_field}
\end{equation}
where $M_{\mathrm{s}}$ is the saturation magnetization and $B_1$ and $B_2$ are the first and second magnetoelastic coupling constants, respectively. The vector $\vec{m} = (m_x, m_y, m_z)$ represents the normalized magnetization and is directly coupled to the corresponding strain tensor components given in Eq.~(\ref{eq:strain_tensor}).
From the LLG equation, the spin-wave propagation characteristics can be described by the dispersion relation developed by Kalinikos and Slavin~\cite{kalinikos1986theory}
\begin{equation}
\label{eq:disprel}
    \omega =\sqrt{(\omega_0+\omega_M\lambda_{\mathrm{ex}}k^{2})(\omega_0+\omega_M\lambda_{\mathrm{ex}}k^{2}+\omega_MF)},
\end{equation}
with
\begin{equation}
\label{eq:disprel_F}
\begin{split}
F &= P + \sin^2(\theta) \bigg( 1 - P \left( 1 + \cos^2(\phi) \right) \\
  &\quad + \frac{\omega_M P(1 - P) \sin^2(\phi)}{\omega_0 + \omega_M \lambda_{\mathrm{ex}} k^2} \bigg)
\end{split}
\end{equation}
and
\begin{equation}
\label{eq:disprel_P}
    P=1-\frac{1-e^{-tk}}{tk}.
\end{equation}
Here, $\omega_{0} = \gamma \mu_0 H_{\mathrm{eff}}$ and $\omega_{M} = \gamma \mu_0 M_{\mathrm{eff}}$ denote the scalar quantities of the effective medium in the spherical coordinate system defined by $\theta$ and $\phi$, where $\gamma$ is the gyromagnetic ratio and $\mu_0$ the vacuum permeability.
The parameter $\lambda_{\mathrm{ex}}$ denotes the exchange constant, and $t$ is the thickness of the magnetic film.
The spin-wave wavenumber is given by $k = 2\pi / \lambda$, where $\lambda$ is the spin-wave wavelength.

For the FVSW configuration, Eq.~(\ref{eq:disprel_F}) simplifies to Eq.~(\ref{eq:disprel_P}) , since the in-plane component vanishes under perfect OOP magnetization ($\theta = 0^{\circ}$).
Accordingly, Eq.~(\ref{eq:disprel}) reduces to
\begin{equation}
\label{eq:disprel_oop}
\omega = \sqrt{(\omega_0 + \omega_M \lambda_{\mathrm{ex}} k^{2}) (\omega_0 + \omega_M \lambda_{\mathrm{ex}} k^{2} + \omega_M P)} .
\end{equation}

Within this formalism, the effects of elastic and plastic deformation, as well as partial amorphization, on spin-wave propagation can be described through Eq.~(\ref{eq:disprel_oop}). The magnetic strain response is captured by $\omega_{0}$. It reflects the accumulation and relaxation of strain through variations in Eq.~(\ref{eq:strain_field}) relative to Eq.~(\ref{eq:strain_tensor}), and consequently in the effective field $H_{\mathrm{eff}}$ as defined in Eq.~(\ref{eq:eff_field}), arising from dislocation accumulation and migration. In contrast, the effects of partial amorphization are accounted for by defining an effective magnetic film thickness, $t = t_{\mathrm{eff}}$, which enters Eq.~(\ref{eq:disprel_oop}) through Eq.~(\ref{eq:disprel_P}).
\begin{figure}[b!] \centering 
\includegraphics[width=\linewidth]{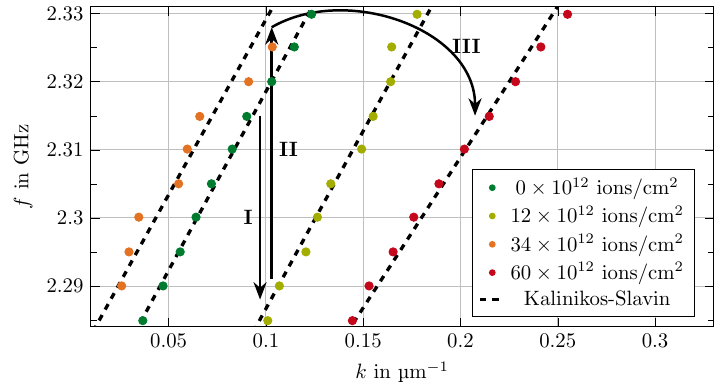}
\caption{Three prominent spin-wave dispersion shifts across the $30$ regions irradiated at $30~\si{\kilo\electronvolt}$. (I) The curve shifts to lower frequencies for implantation doses up to $12\times10^{12}~\si{ions\per\centi\meter\squared}$. (II) For doses increasing up to $34\times10^{12}~\si{ions\per\centi\meter\squared}$, the curve shifts back to higher frequencies. (III) At doses of $60\times10^{12}~\si{ions\per\centi\meter\squared}$ and above, the curve again shifts to lower frequencies and exhibits an additional change in slope.} \label{fig:disp_evol} 
\end{figure}

Accordingly, $M_{\mathrm{eff}}$ and $H_{\mathrm{eff},0}$ are used as fitting parameters only once, by fitting Eq.~(\ref{eq:disprel_oop}) to the dispersion of the pristine film ($H_{\mathrm{mel}}=0$) shown in Fig.~\ref{fig:disp_evol} (dark green markers). The resulting values are then fixed, and subsequent fits to the dispersion in the irradiated regions are used exclusively to extract the dose-dependent variations of $H_{\mathrm{mel}}\neq0$ in Eq.~(\ref{eq:eff_field}).

Under this assumption, we neglect that in the Kalinikos–Slavin formalism magnetoelastic contributions enter the equilibrium effective field, meaning $H_{\mathrm{mel}}$ renormalizes the internal field and modifies the effective magnetization $M_{\mathrm{eff}}$, so both appear in the dispersion relation in a coupled, generally non-separable way. In a joint least-squares fit, all parameters are optimized simultaneously by minimizing weighted residuals, but the strong correlation between $M_{\mathrm{eff}}$ and $H_{\mathrm{mel}}$ prevents a physically meaningful separation of their individual contributions. By assuming negligible variation in $M_{\mathrm{eff}}$ and fixing it, this ambiguity is reduced; although approximate, this approach stabilizes the fitting procedure and enables a more interpretable extraction of the magnetoelastic contribution.

\begin{figure}[h!] \centering 
\includegraphics[width=\linewidth]{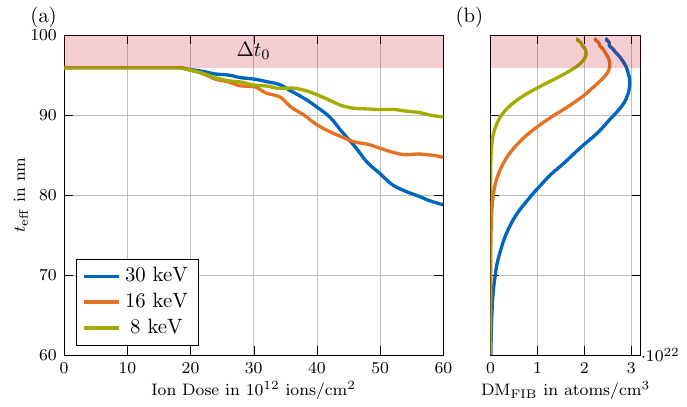}
\caption{Height profiles of the effective film thickness $t_{\mathrm{eff}}$ and simulated damage model $\mathrm{DM_{FIB}}$. (a) AFM height profiles of regions implanted with different ion doses after a $120~\si{\second}$ wet-chemical etching step (etching offset $\Delta t_{0}$), shown for three acceleration voltages. (b) For the highest dose, $60 \times 10^{12}~\si{ions\per\centi\meter\squared}$, the corresponding $\mathrm{DM_{FIB}}$ distribution is displayed, enabling comparison between the measured thickness variations and the simulated ion penetration depth.} \label{fig:height} 
\end{figure}

The dose-dependent $t_{\mathrm{eff}}$ is determined from AFM height measurements performed after wet-chemical etching, as shown in Fig.~\ref{fig:height}(a). In addition to the uniform etching offset $\Delta t_{0}=4.09~\si{\nano\meter}$, irradiated regions exhibit a dose- and energy-dependent height reduction $\Delta t_{\mathrm{FIB}}$, reflecting preferential removal of amorphized material. This thickness reduction is incorporated into the dispersion analysis by $t_{\mathrm{eff}}=100~\si{\nano\meter}-\Delta t_{0}-\Delta t_{\mathrm{FIB}}$.

The trMOKE-derived dispersion data for all irradiated squares at $30~\si{\kilo\electronvolt}$ are fitted using Eq.~(\ref{eq:disprel_oop}), with their corresponding $t_{\mathrm{eff}}$. This analysis reveals systematic, dose-dependent modifications of the spin-wave dispersion. Representative dispersion curves illustrating this evolution are shown in Fig.~\ref{fig:disp_evol}. At low irradiation doses, the dispersion shifts monotonically toward lower frequencies, reflecting a progressive decrease of $H_{\mathrm{mel}}$ associated with strain accumulation (phase I). At intermediate doses, the dispersion trend reverses and shifts toward higher frequencies, indicating a partial relaxation of the local strain due to irradiation-induced dislocation motion (phase II). Upon further increasing the dose, the dispersion shifts downward once more and approaches saturation, consistent with renewed strain buildup combined with near-surface amorphization $\Delta t_{\mathrm{FIB}}$ (phase III).

Complementary SRIM Monte Carlo simulations are employed to determine the depth distribution of irradiation-induced damage. As shown in Fig.~\ref{fig:height}(b), the experimentally observed saturation of the irradiation-induced thickness reduction $\Delta t_{\mathrm{FIB}}$ occurs at depths that are in good agreement with the simulated ion penetration range, hereafter denoted as $\Delta t_{\mathrm{FIB,max}}$. This close correspondence between experiment and simulation suggests that the SRIM simulations provide a reliable approximation of the actual damage profile within the YIG film, consistent with \cite{fodchuk2022effect, hoflich2023roadmap}. Consequently, we assume that the simulated damage distribution can be directly related to the spatial distribution of irradiation-induced deformations and, therefore, to the three-phase scenario. In the following section, this assumption forms the basis for the development of a qualitative, SRIM-based three-phase scenario describing the depth-dependent deformation across the three identified phases.

\begin{figure*} \centering 
\includegraphics[width=\linewidth]{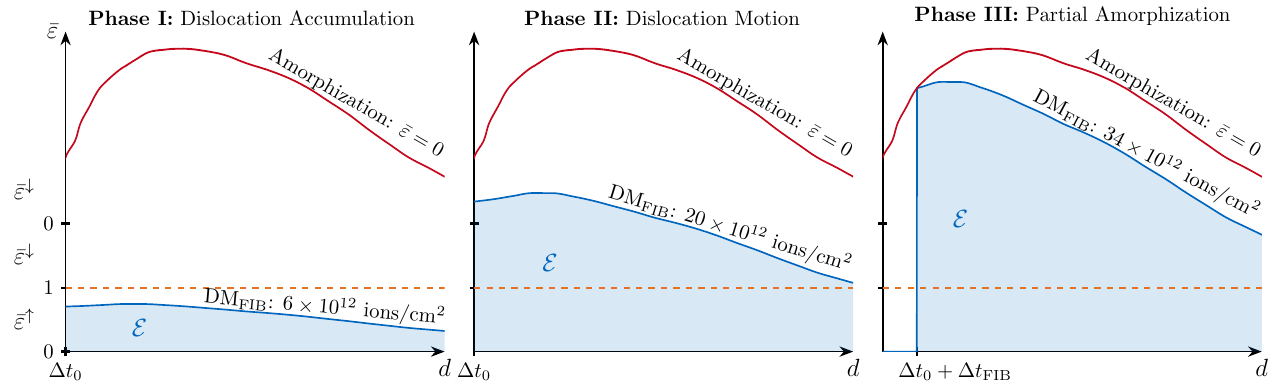}
\caption{Qualitative strain model at $30~\si{\kilo\electronvolt}$ illustrating the three deformation regimes Phase I: Dislocation Accumulation, Phase II: Dislocation Motion, and Phase III: Partial Amorphization. The relative strain $\bar{\varepsilon}(d)$ is shown as a function of depth $d$ for normalized damage models $\mathrm{DM}_{\mathrm{FIB}}$ at ion doses of $6 \times 10^{12}~\si{ions\per\centi\meter\squared}$, $20 \times 10^{12}~\si{ions\per\centi\meter\squared}$, and $34 \times 10^{12}~\si{ions\per\centi\meter\squared}$. The dashed orange line denotes the phase I--II threshold $\mathrm{DM}_{\mathrm{I,II}}$, and the solid red line the depth-dependent amorphization boundary $\mathrm{DM}_{\mathrm{II,III}}$, beyond which $\bar{\varepsilon}=0$. The shaded area indicates the summed relative strain $\mathcal{E}$ used to evaluate the mean relative strain $\langle \bar{\varepsilon} \rangle$.} \label{fig:strain_model} 
\end{figure*}

\subsection{SRIM-Based Modelling of the Three-Phase Scenario}
\label{model}
To interpret the observed dispersion shifts within the framework of the three-phase scenario at a fundamental level, a structural model is required that captures the development of irradiation-induced strain within the crystal. This model provides the physical basis for attributing the fitted dispersion changes to magnetoelastic effects, thereby justifying the extraction of $H_{\mathrm{mel}}$ from the dispersion analysis (Fig.~\ref{fig:disp_evol}). We therefore introduce a qualitative strain model that describes the accumulation, relaxation, and partial breakdown of lattice strain as a function of increasing $\mathrm{Ga}^{+}$ ion dose.

Motivated by the three-phase scenario, the SRIM-based three-phase scenario focuses on the depth-dependent distribution of irradiation-induced damage and the associated onset of dislocation nucleation, as introduced in Sec.~\ref{interpretation}. This distribution is quantified using SRIM simulations. The calculated damage models, denoted $\mathrm{DM}_{\mathrm{FIB}}$, form the basis of the strain description and enable the construction of the qualitative strain profile shown in Fig.~\ref{fig:strain_model} (blue solid line). In this representation, the calculated damage is mapped onto a depth-dependent relative strain profile $\bar{\varepsilon}(d)$.

In the SRIM-based three-phase scenario the three deformation phases are delineated by introducing two characteristic damage boundaries. The first boundary (dashed orange line) marks the end of source hardening, where increasing damage leads to continued strain accumulation, denoted $\bar{\varepsilon}^{\uparrow}$. Beyond this boundary, plastic deformation becomes active and strain relaxation sets in, corresponding to $\bar{\varepsilon}^{\downarrow}$. The second boundary (solid red line) marks the relative strain $\bar{\varepsilon}$ at which strain release is exhausted. Above this boundary, the crystal can no longer accommodate dislocations and transitions into an amorphous state, for which $\bar{\varepsilon}=0$ is assumed.

For an acceleration voltage of $30~\si{\kilo\electronvolt}$, the first phase boundary $\mathrm{DM}_{\mathrm{I,II}}$ is determined phenomenologically. It is calibrated such that the model reproduces the experimentally observed turning point in the dispersion evolution at approximately $12 \times 10^{12}~\si{ions\per\centi\meter\squared}$ (Fig.~\ref{fig:disp_evol}). To this end, $\mathrm{DM}_{\mathrm{FIB}}$ is evaluated at a dose of $8 \times 10^{12}~\si{ions\per\centi\meter\squared}$, yielding a critical damage density of $\mathrm{DM}_{\mathrm{I,II}} = 3.97 \times 10^{21}~\si{atoms\per\centi\meter\cubed}$. Within the model, this choice ensures that the onset of strain relaxation becomes dominant at the experimentally observed turning point. Therefore $\mathrm{DM}_{\mathrm{I,II}}$ is not independently predicted, but serves as a calibrated parameter that aligns the model with the experimental dispersion behavior.

A more rigorous treatment, such as introducing a depth-dependent $\mathrm{DM}_{\mathrm{I,II}}$ or explicitly modeling defect kinetics, would require detailed knowledge of dislocation dynamics and material-specific rate parameters, and is therefore beyond the scope of the present work.

Importantly, this turning point does not correspond to the physical onset of plastic deformation itself. Instead, it marks the regime in which strain relaxation mediated by plastic deformation dominates over further elastic strain accumulation within the effective model. Plastic deformation is assumed to initiate already at lower doses, with the transition between phase I and phase II representing a mixed regime: only the most strongly damaged regions undergo plastic deformation, while less affected depths continue to accumulate strain, consistent with the bell-shaped depth profile of $\mathrm{DM}_{\mathrm{FIB}}$.

In contrast, the second phase boundary $\mathrm{DM}_{\mathrm{II,III}}$, associated with amorphization, is not introduced as a free fitting parameter. Instead, it is determined based on independently measured experimental observables. Specifically, the boundary is obtained by identifying the depth-dependent damage densities required to reproduce the AFM-measured effective thickness $t_{\mathrm{eff}}$ shown in Fig.~\ref{fig:height}(a) (blue line). This procedure anchors the transition to phase III directly to the experimentally observed onset and progression of material removal and amorphization. As a result, $\mathrm{DM}_{\mathrm{II,III}}$ does not correspond to a single critical value but forms a depth-dependent boundary (solid red line in Fig.~\ref{fig:strain_model}), extending from $d=\Delta t_{0}$ to $d=\Delta t_{0}+\Delta t_{\mathrm{FIB}}$, where $\Delta t_{\mathrm{FIB}}$ increases with ion dose.

Since the strain distribution within the crystal is not defined explicitly, we introduce a relative strain metric based on the damage models introduced above. At each depth position $d$, the local strain is defined relative to the phase I–II damage threshold $\mathrm{DM}_{\mathrm{I,II}}$. To this end, we introduce the normalized damage variables
\begin{equation}
x(d) := \frac{\mathrm{DM}_{\mathrm{FIB}}(d)}{\mathrm{DM}_{\mathrm{I,II}}},
\qquad
x_{\mathrm{II,III}}(d) := \frac{\mathrm{DM}_{\mathrm{II,III}}(d)}{\mathrm{DM}_{\mathrm{I,II}}},
\label{eq:damage_var}
\end{equation}
from which the relative strain metric is defined as
\begin{equation}
\bar{\varepsilon}(d)=
\begin{cases}
x(d), & x(d)<1,\\
2-x(d), & 1\le x(d)<x_{\mathrm{II,III}}(d),\\
0, & x(d)\ge x_{\mathrm{II,III}}(d).
\end{cases}
\label{eq:metric}
\end{equation}

This piecewise definition directly maps the local damage state onto the three deformation regimes and captures both the buildup and subsequent release of strain throughout the crystal depth. Moreover, it enforces vanishing strain at depth positions where the local damage exceeds the depth-dependent amorphization boundary $x_{\mathrm{II,III}}(d)$, thereby defining the strain metric in our SRIM-based three-phase scenario. The mean relative strain is therefore evaluated only over those depth positions that remain crystalline,
\begin{equation}
\langle \bar{\varepsilon} \rangle = \frac{1}{n}\mathcal{E},
\qquad
\mathcal{E} := \sum_{i=1}^{n} \bar{\varepsilon}(d_i),
\end{equation}
where the averaging is restricted to depth positions $d_i > \Delta t_{0} + \Delta t_{\mathrm{FIB}}$.

The evolution of $\bar{\varepsilon}(d)$ with increasing ion dose is shown in Fig.~\ref{fig:strain_model} as blue solid lines. For each dose, $\mathrm{DM}_{\mathrm{FIB}}$ is obtained by multiplying the simulated SRIM damage profile by the corresponding ion dose and normalizing it according to Eq.~(\ref{eq:damage_var}). The damage profiles are exemplarily calculated for ion doses of $6 \times 10^{12}~\si{ions\per\centi\meter\squared}$, $20 \times 10^{12}~\si{ions\per\centi\meter\squared}$, and $34 \times 10^{12}~\si{ions\per\centi\meter\squared}$, illustrating the progressive increase of $x(d)$ with dose.

At low ion doses, where $x(d)<1$ throughout the crystalline depth, the relative strain follows $\bar{\varepsilon}(d)=x(d)$ and increases continuously with increasing damage. This regime corresponds to elastic deformation, in which the linear stress–strain response is reflected by the linear increase of $x(d)$ with ion dose. According to the three-phase scenario, dislocations accumulate within the strain–potential well, leading to the continuous strain buildup characteristic of source hardening.

Once the local damage exceeds $\mathrm{DM}_{\mathrm{I,II}}$, such that $x(d)>1$ at specific depth positions, the strain metric enters the second branch of Eq.~(\ref{eq:metric}). This defines phase II, in which $\bar{\varepsilon}(d)=2-x(d)$ decreases with increasing damage, reflecting strain release via plastic deformation. Because $x(d)$ is bounded not only by unity but also by the depth-dependent amorphization threshold $x_{\mathrm{II,III}}(d)$, negative values of $\bar{\varepsilon}$ can be attained when $x_{\mathrm{II,III}}(d)>2$. In this regime, the metric allows for strain states below the initial reference level, which can be interpreted as an improvement of crystalline order compared to the as-fabricated or as-annealed material.

This strain relaxation behavior is inherently nonlinear, as it arises from plastic deformation mediated by dislocation motion. Such nonlinearity cannot be adequately represented by a linearly increasing $\mathrm{DM}_{\mathrm{FIB}}$ alone; instead, it is partially captured through the non-uniform profile of the depth-dependent amorphization boundary $x_{\mathrm{II,III}}(d)$. The curvature of this boundary (Fig.~\ref{fig:strain_model}, red solid line) effectively constrains the number of dislocations that can be released into adjacent strain-potential wells at a given depth $d$, leading to a sequential filling of these wells starting from structural barriers such as the sample surface.

The resulting accumulation of dislocations at such obstacles, along with associated effects such as friction hardening and strain reaccumulation, is not explicitly included in the present model. Consequently, the model does not resolve the full, nonlinear evolution of strain that may arise from competing processes of defect generation, clustering, and annihilation, as described, for example, within kinetic rate-equation frameworks \cite{was2007fundamentals}. Instead, we adopt a linearized approximation of strain accumulation and release, directly followed by partial amorphization initiating at the surface, which implicitly captures the cumulative effects of defect dynamics.

Therefore, we assume that at sufficiently high damage levels, where $x(d) \geq x_{\mathrm{II,III}}(d)$, the corresponding depth positions transition into the amorphous state, and the local relative strain is set to $\bar{\varepsilon}(d)=0$. These amorphized regions are no longer considered part of the active crystalline volume, and the mean relative strain $\langle \bar{\varepsilon} \rangle$ is evaluated only over the remaining effective crystalline thickness $t_{\mathrm{eff}}$.

At greater depths, the material may still exhibit elastic or plastic behavior depending on the local value of $\bar{\varepsilon}(d)$, while the region near $d = \Delta t_{0} + \Delta t_{\mathrm{FIB}}$ progressively transitions into the amorphous state.

\subsection{Justification of the Magnetoelastic Interpretation of the Three-Phase Scenario}
\label{validation}
\begin{figure}[b!]
\centering
\includegraphics[width=\linewidth]{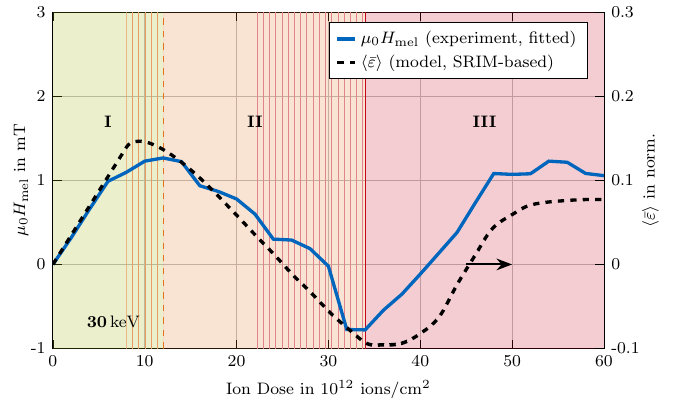}
\caption{Magnetoelastic field variation $\mu_{0}H_{\mathrm{mel}}$ (solid line), extracted from fitting the Kalinikos–Slavin model, as a function of implantation dose. The experimental trend is compared with the simulated mean relative strain $\langle \bar{\varepsilon} \rangle$ (dashed line) modeled by the SRIM-based three-phase scenario at $30~\si{\kilo\electronvolt}$. Both curves exhibit the same characteristic turning points and saturation plateau, reflecting the progression through phases~I–III and their corresponding intermediate regimes.}
\label{fig:strain_30kev}
\end{figure}
Using $H_{\mathrm{mel}}$ in Eq.~(\ref{eq:eff_field}) as the sole fitting parameter in Eq.~(\ref{eq:disprel_oop}), we analyze the dispersion shifts discussed in Sec.~\ref{characterization}. The resulting dose-dependent trend for an ion beam accelerated to $30~\si{\kilo\electronvolt}$ is shown as the solid line in Fig.~\ref{fig:strain_30kev}. Two distinct turning points appear at $12 \times 10^{12}~\si{ions\per\centi\meter\squared}$ and $34 \times 10^{12}~\si{ions\per\centi\meter\squared}$, followed by a saturation plateau at about $50 \times 10^{12}~\si{ions\per\centi\meter\squared}$.

A direct comparison using a fixed scaling ($1~\si{\milli\tesla} \hat{=} 0.1$) between the strain reflected by $\mu_{0} H_{\mathrm{mel}}$ (Fig.~\ref{fig:strain_30kev}, solid line) and the simulated $\langle \bar{\varepsilon} \rangle$ (Fig.~\ref{fig:strain_30kev}, dashed line) reveals a close correspondence between the two quantities. In particular, both curves exhibit the same characteristic turning points and converge to an identical saturation plateau. These signatures correspond directly to phases I–III but also to intermediate regimes in which multiple phases coexist across different material depths. In these intermediate regimes (highlighted by the striped areas in Fig.~\ref{fig:strain_30kev}), individual layers may already enter phase II or phase III, while the overall crystalline structure is still dominated by the preceding phase. The turning points therefore mark the ion doses at which the dominant contribution to the overall strain shifts from phase I to phase II, and later from phase II to phase III. Within the assumptions of the modeled scenario, this parallel evolution supports the interpretation that the two turning points in the dose-dependent spin-wave wavelength observed in Fig.~\ref{fig:dosemap}(b) arise from successive transitions between elastic strain accumulation, plastic strain relaxation, and partial amorphization. The observed agreement supports the plausibility of the proposed three-phase scenario and the corresponding magnetoelastic treatment of ion-induced strain for the $30~\si{\kilo\electronvolt}$, without constituting direct evidence of the underlying microscopic mechanisms.

\begin{figure}[b!]
\centering
\includegraphics[width=\linewidth]{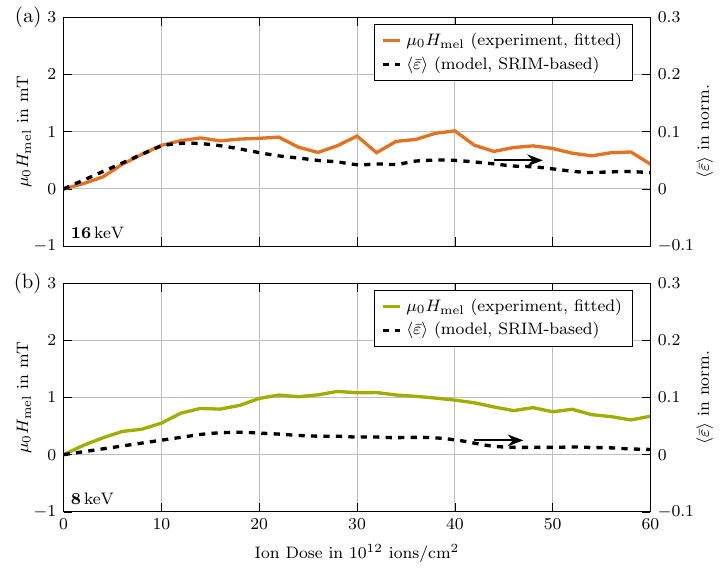}
\caption{Magnetoelastic field variation $\mu_{0}H_{\mathrm{mel}}$ (solid lines), extracted from fitting the Kalinikos–Slavin model, shown as a function of ion implantation dose. The experimental results are compared to the simulated mean relative strain $\langle \bar{\varepsilon} \rangle$ (dashed lines) modeled by the SRIM-based three-phase scenario, using the phase-boundary calibration established at $30~\si{\kilo\electronvolt}$, for acceleration voltages of (a) $16~\si{\kilo\electronvolt}$ and (b) $8~\si{\kilo\electronvolt}$.}
\label{fig:strain_16&8}
\end{figure}

Using the same approach for determining $\mathrm{DM}_{\mathrm{II,III}}$, as established exemplarily for $30~\si{\kilo\electronvolt}$ in Sec.~\ref{model}, we extend our analysis to lower ion-beam acceleration voltages of $16~\si{\kilo\electronvolt}$ and $8~\si{\kilo\electronvolt}$, as shown in Fig.~\ref{fig:strain_16&8}, in order to further assess the robustness of the modeling. The first phase boundary $\mathrm{DM}_{\mathrm{I,II}}$, calibrated for the $30~\si{\kilo\electronvolt}$ case, is kept fixed and directly transferred to the lower energies. In contrast, the second boundary $\mathrm{DM}_{\mathrm{II,III}}$ is re-determined for each acceleration voltage based on the corresponding AFM-measured thickness evolution, thereby maintaining an experimental anchor for the onset of amorphization at each energy.

The extracted magnetoelastic field $\mu_{0} H_{\mathrm{mel}}$ exhibits a consistent magnitude across all acceleration voltages. In contrast, the strain values $\langle \bar{\varepsilon} \rangle$ derived from the SRIM-based three-phase scenario, evaluated using the same fixed scaling for comparison, decrease systematically with decreasing ion acceleration. This results in increasing quantitative deviations between experiment and model, which are most pronounced at $8~\si{\kilo\electronvolt}$ (Fig.~\ref{fig:strain_16&8}(b)). These deviations are primarily attributed to the limited information available on the prestrained state of the film after annealing and its depth distribution prior to irradiation. 

In the present approach, information on the prestrained state is inferred indirectly via the inverse determination of the amorphization boundary using the etch-induced thickness reduction $\Delta t_{\mathrm{FIB}}$, which corresponds to the irradiation-induced thickness change. The maximum thickness reduction $\Delta t_{\mathrm{FIB,max}}$ reflects the maximum penetration depth of the ion beam at a given acceleration voltage. At lower acceleration voltages, the reduced penetration depth and the associated damage cascade result in a smaller $\Delta t_{\mathrm{FIB,max}}$, thereby limiting the depth range over which the prestrained state of the film can be effectively reconstructed.
Consequently, the relative strain profile $\bar{\varepsilon}(d)$ is averaged over an effective thickness $t_{\mathrm{eff}}$, while assuming $\bar{\varepsilon}(d)=0$ for $d>\Delta t_{0}+\Delta t_{\mathrm{FIB,max}}$. This assumption artificially lowers the mean relative strain $\langle \bar{\varepsilon} \rangle$, even though $\bar{\varepsilon}(d)\neq0$ may persist beyond $d>\Delta t_{0}+\Delta t_{\mathrm{FIB,max}}$, where an unknown prestrain is not captured by the model. This reduced depth sensitivity naturally leads to lower apparent $\langle \bar{\varepsilon} \rangle$, owing to the increased contribution of depths where $\bar{\varepsilon}=0$ in the averaging procedure. This effect, in turn, contributes to the observed quantitative deviations at lower ion accelerations. A more accurate quantitative comparison would therefore require a depth-resolved strain profile of the pristine film prior to irradiation. 
Nevertheless, the overall qualitative evolution, and the shape of the strain profile, is consistently reproduced across all acceleration voltages, supporting the robustness of the modeling approach despite the partially phenomenological calibration of $\mathrm{DM}_{\mathrm{I,II}}$ and remaining quantitative deviations in absolute magnitude.

To complete our investigation, we employ micromagnetic simulations to numerically reproduce the experimentally observed wavelength shifts. These simulations do not constitute an independent model for the dispersion evolution. Instead, they serve as a consistency check, demonstrating that the effective strain states inferred from the dispersion analysis and the SRIM-based three-phase scenario are sufficient to reproduce the measured spin-wave behavior within a full magnetoelastic micromagnetic framework. Importantly, all strain-related input parameters used in the simulations are fully determined by the experimentally extracted magnetoelastic field or by the simulated strain model. For this purpose, we examined the strain-tensor components in Eq.~(\ref{eq:strain_tensor}) and their contribution to the magnetoelastic field $\mathbf{H}_{\mathrm{mel}}$ in Eq.~(\ref{eq:strain_field}). Since the actual strain tensor inside the irradiated crystal is unknown, and our dispersion analysis provides only an effective expression for $\mathbf{H}_{\mathrm{mel}}$ in an OOP configuration, further simplifications are required. We therefore assume a small uniaxial deformation in a homogeneous cubic crystal, allowing the strain tensor to be simplified using classical elasticity theory~\cite{landau2012theory}. In this approach, shear components as well as spatial variations of the strain tensor are neglected. Under this assumption, Hooke’s law in three dimensions reduces to
\begin{equation}
  \begin{split}
        \boldsymbol{\varepsilon}&=
        \begin{bmatrix}
            \varepsilon_{xx} & 0 & 0\\
            0 & \varepsilon_{yy} & 0\\
            0 & 0 & \varepsilon_{zz}
        \end{bmatrix}
        =
        \begin{bmatrix}
            -\nu\varepsilon_{zz} & 0 & 0\\
            0 & -\nu\varepsilon_{zz} & 0\\
            0 & 0 & \varepsilon_{zz}
        \end{bmatrix}
        ,     
    \end{split}
  \label{eq:strain_tensor_uniax}
\end{equation}
where a uniaxial strain in the $z$–direction produces transverse strains via the Poisson effect.

Since Eq.~(\ref{eq:strain_tensor_uniax}) does not include shear components, Eq.~(\ref{eq:strain_field}) simplifies to

\begin{equation}
\label{eq:strain_field_oop_hook}
H_{\mathrm{mel}} = -\frac{2}{\mu_0 M_{\mathrm{s}}} B_1 \varepsilon_{zz}.
\end{equation}
which allows us to directly convert the experimentally extracted values of $H_{\mathrm{mel}}$ into the corresponding out-of-plane strain component $\varepsilon_{zz}$.
For the experimental data, the resulting $\varepsilon_{zz}$ values are subsequently inserted into Eq.~(\ref{eq:strain_tensor_uniax}) to determine the transverse strain components $\varepsilon_{xx}$ and $\varepsilon_{yy}$ via the Poisson effect. By contrast, within the SRIM-based three-phase scenario, the simulated $\langle \bar{\varepsilon} \rangle$ is directly mapped onto $H_{\mathrm{mel}}$ using the scaling shown in Fig.~\ref{fig:strain_30kev} and Fig.~\ref{fig:strain_16&8}. The full strain tensor $\boldsymbol{\varepsilon}$ is then constructed analogously using Eq.~(\ref{eq:strain_tensor_uniax}).

Using $M_{\mathrm{s}} = 130~\si{kA\per\meter}$, $B_{1} = 3.48\times10^{5}~\si{J\per\meter\cubed}$, and a Poisson ratio of $\nu = 0.29$, we determine the complete strain tensor $\boldsymbol{\varepsilon}$ for all ion doses within the homogeneously irradiated squares defined in the simulation grid, as described in Sec.~\ref{methods}.

Both the experimentally extracted strain values and those obtained from the SRIM-based three-phase scenario modeled at $30~\si{\kilo\electronvolt}$ already incorporate the effective outcome of the nonlinear lattice response associated with plastic deformation, as encoded in $H_{\mathrm{mel}}$ and $\langle \bar{\varepsilon} \rangle$. Consequently, no additional shear- or plasticity-specific strain components need to be introduced. The resulting strain tensors can therefore be inserted directly into micromagnetic simulations, enabling a quantitative replication of the experimentally observed dose-dependent wavelength shifts shown in Fig.~\ref{fig:dosemap}.

\begin{figure}[t!]
\centering
\includegraphics[width=\linewidth]{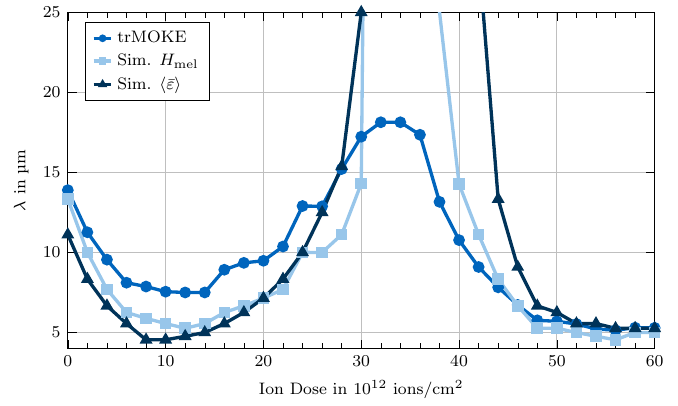}
\caption{Comparison of wavelength shifts as a function of ion dose, extracted from line-wise Fourier transformations of the measured spin-wave profiles at $2.305~\si{\giga\hertz}$ and $8~\si{\mathrm{dBm}}$ under an external field of $\mu_{0}H_{\mathrm{ext}} = 250~\si{\milli\tesla}$. The experimental data (trMOKE) are compared with micromagnetic simulations using two different strain inputs: strain-tensor components derived from the experimentally fitted magnetoelastic field $H_{\mathrm{mel}}$, and the mean relative strain $\langle \bar{\varepsilon} \rangle$ obtained from the SRIM-based three-phase scenario modeled at $30~\si{\kilo\electronvolt}$.}
\label{fig:wavelength_comp}
\end{figure}

Figure~\ref{fig:wavelength_comp} compares the experimentally extracted wavelengths with micromagnetic simulations performed using two different strain inputs: strain-tensor components derived from the experimentally fitted $H_{\mathrm{mel}}$, and the averaged strain $\langle \bar{\varepsilon} \rangle$ obtained from the SRIM-based three-phase scenario modeled at $30~\si{\kilo\electronvolt}$. Owing to the additional assumptions inherent in the SRIM-based three-phase scenario, the corresponding simulated wavelengths exhibit larger quantitative deviations and therefore show reduced agreement with the trMOKE data compared to those based on the experimentally fitted $H_{\mathrm{mel}}$.

Nevertheless, both approaches reproduce the overall wavelength evolution reasonably well, with deviations largely confined to isolated outliers around $32\times10^{12}~\si{ions\per\centi\meter\squared}$. These outliers can be attributed to regimes of reduced spin-wave excitation efficiency, for instance when excitation occurs beyond the ferromagnetic resonance condition or when the resulting wavelengths approach or exceed the finite simulation window.

\section{Discussion}
Despite the simplifications inherent in the analysis, the results provide consistent evidence that FIB-steered spin-wave propagation originates primarily from magnetoelastic effects. The observed turning points in the spin-wave wavelength with increasing ion dose directly correlate with the evolution of a strain-induced magnetoelastic field, accompanied at higher doses by an effective thickness reduction due to partial amorphization of the magnetic layer. Irradiation-induced defects may in principle provide additional scattering channels for spin waves, including two-magnon scattering, defect-induced linewidth broadening, or weak mode conversion. However, within the resolution of the present trMOKE measurements, the extracted propagation damping shows no significant systematic increase with increasing ion dose. Likewise, no clear anomalies in the measured dispersion or damping are observed that would indicate a dominant resonant magnon–phonon contribution. We therefore attribute the observed spin-wave steering primarily to static strain-induced modifications of the magnetic parameters, while possible secondary dynamic contributions cannot be fully excluded.

The experimental findings are well described by a three-phase deformation scenario of crystalline YIG under $\mathrm{Ga}^{+}$ irradiation. This interpretation is supported by a simplified SRIM-based three-phase scenario that captures the dose-dependent damage profile and its correspondence to the three deformation regimes and the observed magnetoelastic behavior. Micromagnetic simulations further corroborate this picture, showing that the combined effects of strain-induced fields and thickness reduction reproduce the measured spin-wave steering.

The distinction between elastic and plastic deformation regimes suggests clear directions for future work. In particular, limiting irradiation to the elastic regime may enable post-annealing recovery, opening pathways toward reversible or reconfigurable strain-based spin-wave control.

However, post-annealing remains technologically challenging in practice. Alternative approaches based on focused laser pulses have recently demonstrated the possibility of locally modifying magnetic properties in YIG thin films via transient melting and recrystallization \cite{levati2025three}. While often described as non-destructive, these processes inherently involve thermally driven structural changes.

In comparison, FIB-based techniques offer exceptional spatial precision, tunability via acceleration voltage, and flexibility in ion species, making them particularly well suited for the realization of highly complex, inverse-designed strain landscapes. A promising direction may therefore lie in combining both approaches: leveraging the nanoscale patterning capabilities of FIB to define precise magnetic index profiles, while employing localized laser-induced re-annealing to enable partial reconfigurability or recovery of the crystalline structure. Such hybrid strategies could provide a pathway toward dynamically adaptable magnonic systems without sacrificing spatial resolution.

Early realizations of magnetic bubble garnet technologies \cite{eschenfelder2012magnetic, wolfe1971modification, jouve1979specific} already demonstrated the versatility of strain-engineered magnetic materials. More recently, magnon straintronics has emerged as a powerful approach to tailor spin-wave properties via strain induced by piezoelectric layers \cite{sadovnikov2018magnon}, enabling reconfigurable spin-wave–based computing architectures and multi-functional devices \cite{grachev2022strain, grachev2024strain}. In particular, recent studies have demonstrated controlled modification of spin-wave band structures, including the formation of tunable magnonic bandgaps and strain-mediated control of spin-wave interference, highlighting the potential of magnetoelastic coupling for the dynamic manipulation of spin-wave propagation.

Strain effects are not limited to conventional magnonic systems but extend naturally to chiral and topological platforms, where broken interfacial symmetry gives rise to the Dzyaloshinskii–Moriya interaction (DMI). In such systems, mechanical deformation has been shown to significantly influence both the strength and directional dependence of DMI, with reports indicating that even moderate strain can alter its sign. In addition, the use of electric fields to generate strain provides a route for reversible and dynamic control of DMI in hybrid structures \cite{gusev2020manipulation, udalov2024electric, li2025modulation}.
Although the present study is restricted to YIG, where inversion symmetry suppresses DMI, the magnetoelastic framework developed here is not material-specific. It can be readily extended to systems with finite DMI, enabling versatile engineering of both conventional and chiral spin-wave properties.

In this context, FIB techniques provide a complementary and uniquely capable platform for strain engineering in the sub-micrometer regime, offering maskless direct-write access to nanoscale and arbitrarily complex strain landscapes. In contrast to piezoelectric approaches, which enable dynamic and reversible control of strain, the present method relies on static, irradiation-induced modifications. This enables precise, spatially programmable control of spin-wave propagation and supports advanced computational functionalities such as frequency demultiplexing \cite{kiechle2023frequency}, while opening pathways toward nonlinear and neuromorphic magnonic devices \cite{papp2021nanoscale}. 

In conclusion, our results establish the magnetoelastic origin of FIB-induced modifications in YIG and thereby provide a foundation for the deterministic, design-oriented use of ion-written strain landscapes. This enables precise spin-wave routing and offers a scalable pathway toward inverse-designed magnonic functionalities.

\section*{Acknowledgment}
This work was funded by the Deutsche Forschungsgemeinschaft (DFG, German Research Foundation) – Project number 514146693.
We further acknowledge the sponsoring support of “Make ideas real. Rohde \& Schwarz @ ZEITlab”, who provided access to the FIB technology.

\bibliographystyle{unsrt}
\bibliography{aapmsamp}
\end{document}